\documentclass[12pt,epsf]{article}
\usepackage{graphicx, amsmath}

\begin{document}

\sloppy
\begin{flushright}{SIT-HEP/TM-33}
\end{flushright}
\vskip 1.5 truecm
\centerline{\large{\bf Generating the curvature perturbation with
instant preheating}}
\vskip .75 truecm
\centerline{\bf Tomohiro Matsuda
\footnote{matsuda@sit.ac.jp}}
\vskip .4 truecm
\centerline {\it Laboratory of Physics, Saitama Institute of
 Technology,}
\centerline {\it Fusaiji, Okabe-machi, Saitama 369-0293, 
Japan}
\vskip 1. truecm
\makeatletter
\@addtoreset{equation}{section}
\def\theequation{\thesection.\arabic{equation}}
\makeatother
\vskip 1. truecm

\begin{abstract}
\hspace*{\parindent}
A new mechanism for generating the curvature perturbation at the end of
 inflaton has been investigated. The dominant contribution to the
 primordial curvature perturbation may be generated during the period of
 instant preheating. The mechanism converts isocurvature perturbation
 related to a light field into curvature perturbation, where the ``light
 field'' is not the inflaton field. This mechanism is important in
 inflationary models where kinetic energy is significant at the end of
 inflaton. We show how one can apply this mechanism to various brane
 inflationary models.
\end{abstract}

\newpage
\section{Introduction}
\hspace*{\parindent}
In the standard scenario of the inflationary Universe, the observed
density perturbations are assumed to be produced by a light inflaton
that rolls down its potential. This ``standard scenario'' has been
investigated by many authors\cite{Books-EU}.
Despite the extensive work, we know that in supersymmetric and
superstring theories there is a serious problem called the
``$\eta$-problem''.
This problem is famous in supersymmetric theories, but it can appear
whenever higher corrections lift the inflaton
potential\cite{Lyth_constraint}.
Although it may be possible to construct some inflationary scenarios
where the flatness of the inflaton potential is protected by symmetry,
it is not always the norm to find a situation where the symmetry appears
naturally and all the required conditions for inflation are satisfied
without any fine-tuning.\footnote{It would be very fascinating if
slow-roll inflation could be embedded in MSSM\cite{MSSM-infla}. }
Recently, a new inflationary paradigm has been
 developed where the conventional slow-roll picture does not play an
 essential role in generating the curvature perturbation.
Along the lines of this ``new inflationary paradigm'', 
we consider in this paper a scenario where isocurvature perturbation of
a light field is converted into the curvature perturbation at the time
of instant preheating when the inflaton kinetic energy is significant
at the end of inflation.
It is significant to note that we will consider a scenario where a
``light field'' is not identified with the inflaton.
The ``light field'' is decoupled from the inflationary dynamics, but it
plays a significant role at the end of inflation.
The idea of a ``new light field'' has already been investigated by many
authors. 
The most famous examples of this would be the curvaton
models\cite{curvaton_1, curvaton_2}. 
In the curvaton models, the origin of the large-scale curvature
perturbation in the Universe is the late-decay of a massive scalar field
that is called the ``curvaton''.
The curvaton paradigm has attracted quite a bit of attention because it
is thought to have obvious advantages.
For example, since the curvaton is independent of the inflaton field,
there was a hope\cite{curvaton_liberate} that the curvaton scenario,
especially in models with a low inflationary scale, could cure serious
fine-tunings associated with the inflation models.
Many attempts have been made to construct realistic models with low
inflationary scale\cite{low_inflation}.
Cosmological defects may play an essential role in low-scale
inflationary scenarios\cite{matsuda_defectinfla}, but the ultimate
solution has not been found yet.
Fast-roll inflation in the standard-model throat could be one of the
successful models following this approach, as we will discuss later in
this paper.
However, Lyth has suggested\cite{Lyth_constraint} that there is a strong
bound for the Hubble parameter during inflation, even when the curvatons
are introduced to a model.
The bound obtained by Lyth\cite{Lyth_constraint} was a critical
parameter in the inflationary model with a low inflation scale, but it
was later suggested by Matsuda\cite{matsuda_curvaton} that the difficulty
could be avoided if an additional inflationary expansion or a phase
transition was present\cite{curvaton_added}.
The idea of a ``light field'' was investigated most recently by
Lyth\cite{delta-N-Lyth}, where an argument was presented that the
density perturbations can be generated
``at the end of inflation'' by the fluctuations of the number of the
e-foldings induced by the fluctuations related to a light scalar field
other than the inflaton.\footnote{See also ref.\cite{alternate}, where 
different models generating a contribution to the curvature perturbation
were proposed. }
The generation of the curvature perturbation is now due to the
fluctuations related to a light field, which is independent of the
inflaton dynamic details, and thus, there could be a similarity between
the curvatons and Lyth's new mechanism in the usage of a light field.
However, there is an important advantage in the new mechanism in that 
unlike the curvatons, one does not have to worry about the stringent
conditions that come from the requirement of the successful late-time
dominance and reheating.
Using this new idea, we studied the generation of the curvature
perturbation without using the slow-roll
approximations\cite{matsuda_elliptic,matsuda_ellip2}. 
The situation we considered is very common in the brane inflationary
models.\footnote{Riotto and Lyth made another useful discussions of this
point\cite{Riotto-Lyth}.} 
We considered a brane inflationary model where a light field appears
``at a distance'' from the moving brane.
This idea fits in well with the generic requirements of the models with
``throat'' inflaton.
To be more precise, we demonstrated that the correction lifting the
moduli space in the bulk is not a serious problem if there is a symmetry
enhancement ``at the tip'' of the KS throat\cite{KS-throat}.
However, it is still not clear if the mechanism can work in a model
where the inflationary brane has (ultra) relativistic velocity, such as
in the cases of DBI inflation or trapped inflaton.
Naively using Lyth's mechanism in these cases is inappropriate,
since the hybrid-type potential does not play a significant role. 
Moreover, it is unclear what happens if the expectation value of
the light field (i.e. the impact parameter) is so large that the
inflaton cannot directly hit the waterfall region of the hybrid-type
potential. 
If the brane motion is ultrarelativistic, or the kinetic energy of the
inflaton is significant, the so-called ``instant preheating'' should be
important in the analysis. Our new mechanism for generating the
curvature perturbation is very useful because this mechanism
 can be used to discuss
the generation of the curvature perturbation in inflationary models
``without'' branes. 
Thus, what we will consider in this paper is a generic
mechanism for generating the curvature perturbation at the end of
inflaton that works with instant preheating and a light
field. 
Construction of a successful brane inflationary model is one of
our motivations, but there is no reason we have to stick to the brane
Universe, as our mechanism is applicable to generic inflationary
models. 

In our scenario, as we have discussed previously\cite{matsuda_ellip2},
it is better to lower the scale of inflation so that one can obtain a
large number of e-foldings without slow-roll. 
On the other hand, the
requirements from the generation of the curvature perturbation place a
lower bound on the inflationary scale. 
This problem can be solved if
there is a preceding short period of inflationary expansion, the
solution of which is similar to the mechanism discussed by us for the
curvatons\cite{matsuda_curvaton}.
Assuming that the fluctuation of a light field is generated during the
first inflationary expansion with the potential $V_1\sim M_1^4$, and also
that the instant preheating occurs at the end of secondary inflation
with the potential $V_2\sim M_2^4$, the curvature perturbation is
enhanced by a factor of $O(M_2^2/M_1^2)$, which makes it possible to
have low inflationary scales in both the first and the second
inflationary epochs.

Aside from the generation of the curvature perturbation, one might think
that the dangerous overproduction of the unwanted cosmological relics
might be critical. For example, since the enhanced isometries are known
to make the Kaluza-Klein modes completely stable, these ``stable'' relics
may put an unavoidable lower bound on the inflationary scale in the
brane Universe\cite{KK-string_Kofman}. 
Although this problem was already discussed in a similar
context\cite{matsuda_ellip2}, it should be better to revisit this
problem again in this paper to complete our analyses. 

\section{Generating the curvature perturbation with instant preheating}
\label{Model_overview}
We consider a multi-field inflationary model where the background
inflaton fields $\phi_{i}(t)$ evolve according to the system of coupled
differential equations
\begin{equation}
\ddot{\phi_i}+3H \dot{\phi_i}+\frac{\partial V}{\partial \phi_i}=0, 
\, \, \, \, \, \, i=1,...,n
\end{equation}
and 
\begin{equation}
H^2 = \frac{8\pi}{3M_p^2}\left[
\sum_i \frac{\dot{\phi}_i^2}{2} + V \right].
\end{equation}
Without loosing general applicability,
 we can discuss our mechanism
with two fields, $\phi_1$ and $\phi_2$, where $\phi_1$ is the
conventional inflaton field and $\phi_2$ is the ``light field''.
The potential $V(\phi_1, \phi_2)$ is characterized by a hierarchy
between the masses of the fields, and can be modeled by 
\begin{equation}
V(\phi_1, \phi_2)= \frac{m_1^2}{2}\phi_1^2 + \frac{m_2^2}{2}\phi_2^2,
\end{equation}
where $m_1 \sim O(H)$ and $m_2 \ll m_1$.\footnote{A different approach
has been given by Kolb et.al.\cite{SSB-curvaton}, who discussed
 a new general mechanism to generate curvature perturbation after the
 end of the slow-roll phase of inflation.
This model is based on the simple assumption that the potential driving
inflation is characterized by an underlying global symmetry which is
slightly broken. See also table\ref{Fig:table}, where the
mechanisms for generating the curvature perturbation at the end 
  of inflation are categorized in four groups.
  We are mainly concerned about the $\eta$-problem and did not use the
  slow-roll approximations.} 
\begin{table}[h]
\begin{tabular}{|c|c|c|}
\hline
& Inflation ends with waterfall & Inflation ends with preheating\\
\hline
 $m_1 \gg m_2$ & D. H. Lyth\cite{delta-N-Lyth} & This paper\\
\hline
 $m_1 \sim m_2$ & T. Matsuda\cite{matsuda_elliptic} & W. Kolb, A. Riotto and 
A. Vallinotto\cite{SSB-curvaton}\\
\hline
\end{tabular}
\caption{Mechanisms for generating the curvature perturbation at the end of inflation }  
\label{Fig:table}
\end{table}

We consider the instant preheating model\cite{inst_preheating}
as the process through which the inflaton decays into lighter particles.
The typical coupling to the preheat field $\chi$ is written as
\begin{equation}
{\cal L}=\frac{g^2}{2}(\phi_1^2+\phi_2^2) \chi^2,
\end{equation}
which gives a mass $m_{\chi} = g \sqrt{\phi_1^2+\phi_2^2}$ 
to the preheat field.
Applying the result obtained in ref.\cite{inst_preheating}, 
the comoving number density $n_{\chi}$ of the preheat field $\chi$
produced during the first half-oscillation of $\phi_1$
becomes\footnote{See appendix A for more details.}
\begin{equation}
\label{n_chi}
n_\chi = \frac{(g|\dot{\phi_1}(t_*)|)^{3/2}}{8\pi^3}
\exp\left[-\frac{\pi g |\phi_2(t_*)|^2}{|\dot{\phi_1}(t_*)|}
\right],
\end{equation}
where $t_*$ is the time when the inflaton $\phi_1$ reaches its minimum
potential at $\phi_1=0$ and where the light field $\phi_2$ may still
have an expectation value $\phi_2(t_*)\ne 0$.
We used $\dot{\phi}_2 =0$, $\delta
\dot{\phi}_2 =0$ and $\phi_1(t_*)=0$ to derive eq.(\ref{n_chi}).
To obtain an estimate of the curvature perturbation through
eq.(\ref{n_chi}), we need to write down an expression for $\delta
n_\chi/n_\chi$;
\begin{equation}
\label{1storder}
\frac{\delta n_\chi}{n_\chi} \simeq
-\frac{2\pi g |\phi_2(t_*)|^2}{|\dot{\phi}_1(t_*)|} 
\frac{|\delta \phi_2(t_*)|}{|\phi_2(t_*)|},
\end{equation}
where it is assumed that $|\delta\phi_2(t_*)| \ll |\phi_2(t_*)|$.
To determine the curvature perturbation produced during the decay
process of the preheat field $\chi$, it is sufficient to note that the
generated energy density is proportional to the comoving number density
$n_\chi$. 
Assuming a smooth decay process of the preheat field, the curvature
perturbation $\zeta$ generated during the instant preheating is 
\begin{equation}
\zeta \simeq \alpha \frac{\delta n_\chi}{n_\chi},
\end{equation}
where $\alpha$ is a constant whose numerical value depends on the
redshif of the particle produced.
Since the field $\phi_2$ is approximately massless during inflation,
the value of the fluctuation is given by $\delta \phi_2 \sim H_I$.
In the simplest case of a single-stage inflationary model,
the curvature perturbation $\zeta$ is approximately 
\begin{equation}
\label{xi_1}
\zeta \simeq \frac{\alpha g |\phi_2(t_*)|^2}{m |\phi_1(t_i)|}
\frac{H_I}{|\phi_2(t_*)|} \sim 
\left|\frac{\phi_2(t_*)}{\phi_1(t_i)}\right| \left(
\frac{H_I}{m}\right),
\end{equation}
where $t_i$ is the time when the inflaton $\phi_1$ starts fast-rolling,
and $m_1^2 |\phi_1(t_i)|^2 \simeq |\dot{\phi}_1(t_*)|^2$ is used to derive
the equation.
Since the field $\phi_2$ is very light, it is possible to have 
$\phi_2(t_i)\simeq \phi_2(t_*)$.
If there is the ``$\eta$-problem'', the natural scale of the mass $m$
would be $m^2 \ge O(H_I^2)$.\footnote{The mass of a light field $\phi_2$
is protected by symmetry, as we will discuss later.}
Moreover, since we are considering a case where kinetic energy is
significant at the time of preheating, 
\begin{equation}
\label{natu_con}
\dot{\phi}_1(t_*)^2 \sim m^2 |\phi_1(t_i)|^2 \sim H_I^2 M_p^2
\end{equation}
is a natural consequence.
We must also consider the condition for the efficient production of the
preheat field $\chi$, which is written as
\begin{equation}
m_\chi^2 \simeq g|\phi_2(t_*)|^2 < \dot{\phi}_1(t_*).
\end{equation}
As we will discuss later in Sect.\ref{Brane-inf}, the effective mass of
the preheat field $\chi$ is suppressed by a Lorentz factor in DBI
inflation, where the motion of the inflation becomes (ultra) relativistic.
We postpone the analyses of the ultra-relativistic motion.

Putting these conditions into (\ref{xi_1}), we obtained an order
estimation $\zeta < \sqrt{H_I/M_p}$.
The above condition is not favorable for the brane Universe
with a low-scale.
Is it possible to remove this condition without
introducing fine-tunings?
As we discussed in Sec.1, a similar problem has already been 
discussed in the curvaton models.
A lower bound for the inflationary scale appeared in the curvaton
model and was discussed by Lyth\cite{Lyth_constraint}, and a solution 
was suggested by us\cite{matsuda_curvaton}.
Following the solution\cite{matsuda_curvaton}, we have attempted 
to introduce another inflationary stage that has the Hubble constant
$H_I'>H_I$. 
Denoting the typical scale of the inflationary potential by $M$ and $M'$
for each inflation, and assuming that the relevant perturbation of the
light field $\phi_2$ is generated during the preceding inflation with 
the Hubble constant $H_I'$, we obtained in a straightforward manner 
the order estimation of the curvature perturbation
\begin{equation}
\label{zeta-1}
\zeta \sim 
\left|\frac{\phi_2(t_*)}{\phi_1(t_i)}\right| \left(
\frac{M^{'2}}{M^2}\right),
\end{equation}
where $m \simeq H_I$ is assumed.
The factor $(M'/M)^2$ helps the generation of the curvature perturbation
in low-scale inflationary models.

\section{Brane Universe}
\label{Brane-inf}
The origin of a ``light field'' could be the isometries in the KS-throat,
which is a useful idea that we have discussed in Sect.1.
At least in the present model of the KS throat there are isometries 
at the tip of the throat, where the target brane is waiting
for the inflationary (moving) brane.

If brane inflation is induced by a hybrid-type brane-antibrane
potential, and also the kinetic energy of the inflationary brane is not
significant, inflation ends with the waterfall of the conventional
hybrid-type potential.
In this case, one can use Lyth's mechanism for generating the
curvature perturbation at the end of inflation.
Since the waterfall is essential to this model, one has to introduce
mild fine-tunings in the initial conditions.
To be more precise, the ``light field'' should not have its expectation
value larger than the string scale at the tip, which corresponds to the
radius of the ``waterfall area'' of the hybrid-type potential.
The situation is shown schematically in Fig.\ref{Fig:basic}. 
Moreover, the motion of the ``inflaton field'' must not be
ultra-relativistic so that the hybrid-type potential can play an essential
role at the end of inflation.

On the other hand, we know it is possible to construct inflationary
models in the brane Universe, which do not satisfy the above
criteria.
For example, as discussed by Alishahiha and Tong\cite{Tong-DBI} the
ultra-relativistic motion of the inflationary brane may play a 
crucial role
in some inflationary models in the brane Universe.
Another kind of inflaton was discussed by Kofman et.al.\cite{beauty_is},
in which trapping of a moving brane may induce weak
inflationary expansion. 
It is important to note that the fast motion of an
inflationary brane does play an essential role in these models, but the
hybrid-type potential does not.
Then, we have a natural question as to whether 
it is possible to generate the
curvature perturbation at the end of inflaton in these alternative
inflation models. 
In this section we give an answer to the above question.

Related to the brane-brane or brane-antibrane collision at the end of
inflation, two points should be clarified.
One is that at instant preheating the mass of the $\chi$ field
might reach a negative value if $\chi$ is the tachyon that appears in a
brane-antibrane collision.
If so, the collision does not correspond to the instant preheating
scenario we referred to in Sect.2.
The tachyon mass becomes negative when the distance between brane and 
anti-brane becomes shorter than the critical length, hence the 
preheating must be calculated with the negative tachyon mass if instant
preheating occurs ``within'' the region $\phi_2 < M_*$, where $M_*$
denotes the string scale at which the mass of the $\chi$ field becomes
negative.
Hence, the underlying condition for the
standard instant preheating scenario is 
(1) the collision is brane-brane(not brane-antibrane), or
(2) the minimum length between branes at the preheating is longer than 
the critical length if the collision is brane-antibrane. 
Once the preheat field is produced at the collision, they will soon 
obtain huge mass.
The other point is that since we are using light open strings on the branes
as the preheat field, the DBI action should be the starting point to see
how the $\chi$ field is generated in instant preheating.
More specifically, the concern is whether the higher derivative terms
in the DBI 
kinetic term may or may not play distinguishable role in instant
preheating.
Although this issue seems interesting, we will not deal
with this difficult issue in this paper.
We will make a simplification to capture the
characteristics of the brane collision\cite{beauty_is,ultra-motion}.

\subsection{Trapped inflation}
Consider the theory of a real scalar field $\phi_1$ with the effective
potential $V_1 \simeq m^2 \phi_1^2 /2$, where
the mass is larger than the Hubble constant, and
thus the inflaton $\phi_1$ falls rapidly to its minimum\cite{beauty_is}.
We will assume that $\phi_1$ falls from its initial value 
$\phi_{1,0}=\phi_{1*}(1+\alpha)$ with vanishing initial speed and 
gives some boson (preheat field) a mass
\begin{equation}
m_\chi^2 \simeq g^2 ((\phi_1-\phi_{1*})^2 + \phi_2^2),
\end{equation}
 where $\phi_2$ is a light field protected by symmetry.
Then, as $\phi_1$ passes $\phi_{1*}$, it creates $\chi$ particles with
number density $n_\chi \simeq \frac{(gv)^{3/2}}{8\pi^3}\exp\left(-\pi g
\phi_2^2/v \right)$, where $v$ denotes the velocity of the field
$\phi_1$.
After a short time these $\chi$ particles become nonrelativistic,
and induce effective potential
\begin{equation}
V_{eff} \simeq \frac{1}{2}m^2 \phi_1^2 + 
g n_\chi \sqrt{(\phi_1-\phi_{1*})^2 + \phi_2^2},
\end{equation}
where $\sqrt{(\phi_1-\phi_{1*})^2 + \phi_2^2} \simeq (\phi_1-\phi_{1*})$
 is a natural assumption.
Subsequent expansion of the Universe dilutes the density of $\chi$
particles as $n_\chi \propto a^{-3}$, which eventually reduces the
$n_\chi$-dependent correction to
the effective potential.
Then, the inflaton $\phi_1$ starts moving down again.
Let us calculate the number of e-foldings assuming (for simplicity)
$g\sim \alpha \sim 1$.
The inflaton is trapped at $\phi_1 \sim \phi_{1*}$ until the scale
factor of the Universe $a$ grows so that the gradient of the potential
changes its sign at
\begin{equation}
a^{-3} v^{3/2}\exp\left(-
\phi_2^2/v \right) \sim m^2\phi_1.
\end{equation}
The number of e-foldings occuring  during this ``trapping'' period
is\cite{beauty_is}  
\begin{equation}
N_e \sim \frac{1}{6}\ln(\frac{\phi_1}{m})
- \frac{1}{3}\left(\frac{\phi_2^2}{v} \right),
\end{equation}
where the original assumption in 
ref.\cite{beauty_is} was that $\phi_2 \simeq 0$ .

Let us consider a fluctuation of a light field $\phi_2$.
The fluctuation of the number of e-foldings is induced by
 the fluctuation related to the light field $\phi_2$.
Assuming $|\phi_2| \gg \delta \phi_2$, we obtained 
\begin{equation}
\label{zeta-2}
\zeta\simeq \delta N \sim \frac{\phi_2}{v}\delta \phi_2,
\end{equation}
where the curvature perturbation with a given wavenumber $k$ of
cosmological interest is $\zeta(k) = \delta N(k)$.

\subsection{DBI inflation}

The ``standard'' adiabatic perturbation generated in
DBI inflation has been calculated by Alishahiha et. al.\cite{Tong-DBI}.
The result is 
\begin{equation}
\label{Zeta-DBI}
\zeta_{DBI} \simeq \frac{\sqrt{g_s} H^2}{\dot{\phi_1}(t_i)},
\end{equation}
where $g_s$ is the squared gauge coupling of the corresponding gauge
dynamics. 
The basic idea of the ``D-cceleration'' is given by the 
evolution of $\phi_1(t)$ which is fixed by the first order Friedman
equation. 
The above result is obtained when the dynamics of the probe D3-brane is
captured by the Dirac-Born-Infeld action coupled to
gravity\cite{Tong-DBI},
\begin{eqnarray}
S&=&\int \frac{1}{2}M_p^2 \sqrt{-g}{\cal R}+ {\cal L}_{eff}+ ...\nonumber\\
 {\cal L}_{eff} &=& -\frac{1}{g_s}\sqrt{-g}\left[
f(\phi)^{-1} \left(1+f(\phi)g^{\mu\nu}\partial_{\mu}\phi
\partial_{\nu}\phi\right)^{1/2}+V(\phi)
\right].
\end{eqnarray}
Alishahiha et. al. suggested in Ref.\cite{Tong-DBI} that $\dot{\phi_1}$
is then given by the formula\footnote{See appendix B in Ref.\cite{Tong-DBI}}
\begin{equation}
 \dot{\phi}_1 = -2g_s M_p^2 \frac{H'(\phi_1)}{\gamma(\phi_1)} \simeq -f^{-1/2},
\end{equation}
where $\gamma(\phi_1)=\sqrt{1+4g_s^2M_p^4 f(\phi_1)
H'(\phi_1)^2}$.
The important point is that for $\gamma \gg 1$ the ``speed limit''
dominates the dynamics of the scalar field $\phi_1$ and leads to
an interesting cosmological scenario 
which differs from the usual slow-roll scenarios.
Here $f(\phi_1)$ is the warp factor of the AdS-like throat which is given
by
\begin{equation}
f(\phi_1)\simeq \frac{\lambda}{\phi_1^4}.
\end{equation}

Now let us extend our result  to apply to a DBI
inflationary model discussed in Ref.\cite{Tong-DBI}.
Here we assume that inflation ends with instant preheating at
$\phi_1(t_*) >0$, where a static brane is waiting for a inflationary
brane.
In backgrounds with $\dot{\phi}_1\ne 0$, there are velocity-dependent
corrections to the effective action, which shifts the effective mass
\begin{equation}
m_\chi|_{DBI} \propto \frac{m_\chi}{\gamma},
\end{equation}
where $m_\chi|_{DBI}$ is the mass of a string stretched from an
 inflationary 
 brane to another brane waiting at the tip\cite{Tong-DBI}.
Hence $\gamma$ appears in our result as
\begin{equation}
\label{zeta-PR-DBI}
\zeta_{PR-DBI} \simeq \alpha \frac{\delta n_\chi}{n_\chi}\simeq 
\alpha \frac{|\phi_2||\delta \phi_2(t_*)|}{\gamma^2\dot{\phi_1}(t_*)},
\end{equation}
where we assumed that $\gamma$ is a slowly-varying function of $\phi_1$.
Here we can express $\gamma$ as\cite{Tong-DBI} 
\begin{equation}
\gamma^2 \simeq \frac{4g_s^2}{\epsilon_{DBI}^2}\frac{M_p^4}{\phi_1^4}
\simeq \frac{4g_s \lambda}{3}\frac{M_p^2 m_1^2}{\phi_1^4},
\end{equation}
where $\epsilon_{DBI}\simeq \sqrt{3g_s/\lambda}M_p/m_1$ is a ``slow-roll''
parameter for DBI inflation\cite{Tong-DBI}. 
Comparing Eq.(\ref{zeta-PR-DBI}) and (\ref{Zeta-DBI}), we find that
the condition $\zeta_{DBI}/\zeta_{PR-DBI} \gg 1$ that is needed for the
``standard'' fluctuation $\zeta_{DBI}$ to dominate over $\zeta_{PR}$ is
written as 
\begin{eqnarray}
\frac{\delta \phi_2}{\phi_2}&\gg& \frac{\alpha}{\sqrt{g_s}\gamma^2(t_*)} 
\frac{\dot{\phi}_1(t_i)}{\dot{\phi}_1(t_*)}\nonumber\\
&\simeq & \frac{\alpha}{\sqrt{g_s}} 
\frac{\phi_1(t_i)^2}{\gamma^2(t_*)\phi_1(t_*)^2}\nonumber\\
&\simeq& \frac{\phi_1(t_i)^2 \phi_1(t_*)^2}{g_s^{3/2}\lambda M_p^2 m_1^2},
\end{eqnarray}
where $\delta \phi_2 \simeq H /2\pi$ is considered.
Following Ref.\cite{Tong-DBI} we suppose that
$\phi_1(t_i)\sim \sqrt{g_s} M_p$, $\phi_1(t_*)\sim m_1$, and
$\lambda \sim g_s \times 10^{10}$.
Then the above condition becomes
\begin{equation}
\frac{\delta \phi_2}{\phi_2}\gg g_s^{-1/2} \lambda^{-1} \sim
g_s^{-3/2}\times 10^{-10}.
\end{equation}
The crucial factor of $10^{-10}$ comes from a normalization
of the curvature perturbation obtained in the original DBI inflationary
model
\begin{equation}
\zeta_{DBI} \simeq g_s^{1/2}/(\epsilon_{DBI}^2 \lambda^{1/2}) \simeq
\lambda^{1/2} m^2/(g_s^{1/2} M_p^2)\simeq 10^{-5} 
\end{equation}
and the assumption $m\simeq 10^{13}$GeV that was made in
Ref.\cite{Tong-DBI}. 
In terms of the gravity dual, $\lambda$ is given by
$\lambda \sim (R/l_s)^4$, which is determined by the radius of the ADS
space $R$ and the fundamental string length $l_s$.
The above condition suggests that the ``standard'' perturbation
$\zeta_{DBI}$ appears exclusively in the spectrum.
Moreover, one can see that the dimensionless parameter $\lambda$ and the
mass scale $m$ is 
constrained by the slow-roll condition $\epsilon_{DBI}\le 1$, which is
needed to obtain sufficient number of e-foldings during D-cceleration,
\begin{equation}
N_e =\epsilon^{-1}_{DBI} log \left(\frac{\phi_1(t_i)}{\phi_1(t_*)}\right).
\end{equation}
Actually, the slow-roll condition $\epsilon_{DBI}<1$ require
$g_s/\lambda < (m/M_p)^2$ while the amplitude of tensor modes requires
$m^2/M_p^2 < 10^{-10}$, which leads to the same conclusion without using
the normalization of the primordial curvature perturbation.
Hence our conclusion is that $\zeta_{PR-DBI}$ is generically much
smaller than $\zeta_{DBI}$ if sufficient number of e-foldings is
generated during DBI inflation.

On the other hand, there might be a possibility that the required 
number of e-foldings is generated during precedent inflation in the bulk,
while reheating is induced by a relativistic collision at the tip.
Of course this scenario is not a DBI inflationary model since in this case
inflation is not due to D-cceleration.
One may think that this kind of scenario is more generic than the
original DBI inflationary model.
In this case, to obtain $\zeta_{PR-DBI}\simeq 10^{-5}$ one must choose
$\zeta_{PR-DBI}\simeq \phi_2 m_1/(\lambda^{1/2} M_p^2) \sim 10^{-5}$ for 
$\dot{\phi_1}(t_*)\simeq f^{-1} \simeq m_1^2/\lambda^{1/2}$  
and $\delta \phi_2 \simeq m_1$\cite{Tong-DBI}, which leads to a
condition $\phi_2 > \lambda^{1/2} M_p$.
Hence the condition $\zeta_{PR-DBI}\simeq 10^{-5}$ requires large 
$\phi_2$ which is larger than the largest allowed value of $\phi_2$
($\sim g_s^{1/2}M_p$)  in Ref.\cite{Tong-DBI}.
The value $\phi_2 > \lambda^{1/2} M_p$ is also larger than the
largest impact parameter allowed by instant preheating.

\begin{figure}[h]
 \begin{center}
\begin{picture}(440,220)(0,0)
\resizebox{13cm}{!}{\includegraphics{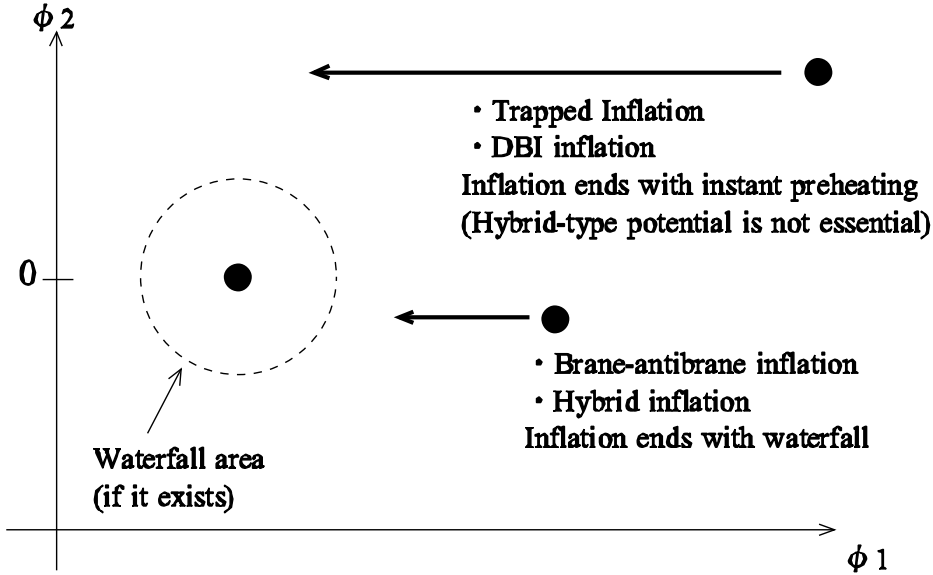}} 
\end{picture}
\caption{Brane-antibrane inflation ends with reheating
  induced by the waterfall field(tachyon), if kinetic energy is not
  significant. 
In this case, the mechanism advocated by Lyth\cite{delta-N-Lyth} may
 play a significant role in generating the curvature perturbation at the
  end of inflation, where the fluctuation $\delta \phi_2$ is converted
  into $\delta N$.  
On the other hand, brane-brane collision or brane-antibrane collision
  with significant kinetic energy cannot be explained by
 conventional waterfall. }  
\label{Fig:basic}
 \end{center}
\end{figure}

Hence, our conclusion in this section is that the ``standard''
calculation in DBI inflation cannot be accommodated by the curvature
perturbations generated at the end of inflation.
This result is true once the D-cceleration becomes crucial during
or even after inflation. 
On the other hand, the situation becomes marginal if brane collision is
not ultra-relativistic.
In this sense, the previous attempts\cite{delta-N-Lyth, matsuda_elliptic, SSB-curvaton}
 and our present
model may complement each other to give the paradigm of ``generating the
curvature perturbation at the end of inflation'', as is shown in
Fig.\ref{Fig:basic}. 

\section{Cosmological relics}
\label{KK-string}
Because of the $\eta$-problem, our new scenario favors low inflationary
scale\cite{matsuda_ellip2}.
The question as to whether there are further constraints coming from
other cosmological considerations with low inflationary scale requires
an answer.
Although there is general knowledge of the various constraints in this
direction, the constraints are sometimes based on some specific
conditions that are highly model-dependent.
A generic condition that places a significant bound on the scale of the
usual hybrid-type inflationary model is discussed by
Lyth\cite{TeV-hybrid}, who suggested that 
the inflaton-waterfall-field coupling induces a one-loop
contribution to the inflaton potential and ruins the slow-roll
approximation as well as the generation of the structure of the Universe. 
Therefore, one cannot have a successful slow-roll inflationary model of
a hybrid-type potential with a scale lower than $10^9$ GeV.
Our model does not suffer from this condition. 
Besides the above condition, other
conditions coming from the fact that the typical mass scale of the brane
binding the maximum value of a scalar field on the brane must be
considered.
If the fundamental scale is as low as $M \sim$O(TeV), the
inflation field must be a bulk field, and thus one cannot exclude the
serious constraint coming from the decaying Kaluza-Klein
mode\cite{matsuda_KK},
since such bulk fields always couple to the Kaluza-Klein states.
The crucial point is that the constraint appears even if one discards
the slow-roll approximations.
In relation to a brane inflationary model, it has been
suggested\cite{KK-string_Kofman} that the constraint coming from the 
Kaluza-Klein mode becomes much more 
serious in brane inflationary models with some isometries in the
compactified space, since the isometries stabilizes the unwanted
KK-relics.
This condition is quite significant in our scenario and
 excludes low-scale inflation with $M < 10^{12}$GeV.
More
recently, however, there has been discussion\cite{warped-reaheating}
that in studying a more detailed thermodynamic 
evolution of the heating process, especially that of the KK particles,
many qualitatively different results compared to the original result
can be found. 
Following the new result\cite{warped-reaheating},
we may consider throat inflation in the standard-model throat with the
inflationary scale $\sim$O(TeV), even if an isometry at the tip of the
throat plays the crucial role in our scenario.
One may also worry about dangerous cosmological ``defects'' that may be
produced during brane annihilation. 
It is known that cosmic strings can put an
upper bound on the inflationary scale of the last
inflation\cite{string_relic}, but the bound is not significant in our
scenario.
Besides the cosmic strings, monopoles and walls may be produced as a
consequence of brane creation or brane deformation that may occur during
or after the reheating epoch\cite{matsuda_brane_monopolesandwalls}. 
It is known that these defects might put a strict upper bound on the
inflationary scale, if there is a non-trivial structure in the
compactified space. 
A natural solution to the domain wall problem in a typical supergravity
model is discussed by us\cite{matsuda_wall}, where the required
magnitude of the gap in the quasi-degenerated vacua is induced by $W_0$
in the superpotential.
The mechanism discussed\cite{matsuda_wall} is quite 
natural since the constant term $W_0$ is the one that 
appears to cancel the cosmological constant.
The bound obtained in ref.\cite{matsuda_brane_relics} is severe, but
it is also known that the structure that is required
to produce the dangerous defect configuration of the brane does not
appear in the known example of the KS throat. 
Of course, it is not clearly understood whether it is possible to obtain
the complete Standard Model(SM) in the known example of the KS
throat. Thus, it is fair to conclude here that this problem is
unsolved and requires further investigation together with the
construction of the complete set of the SM model in the brane Universe. 

In lowering the inflationary scale it might be thought that it would be
difficult to obtain enough baryon number asymmetry of the Universe
(BAU), since the requirement of the proton stability puts a strict bound
on the baryon-number-violating interactions. 
This speculation is indeed true. 
The old scenario of the baryogenesis with a decaying heavy
particle cannot work if the fundamental scale goes as low as the O
(TeV) scale\cite{low-x-decay-baryo}.
One can solve this problem by introducing cosmological defects that
enhance the breaking of the baryon number conservation in the core
\cite{low-x-decay-baryo_matsuda}.
There is also a problem in the scenario of Affleck-Dine
baryogenesis\cite{AD}, since the expectation value of a field on a brane
cannot become much larger than the typical mass scale of the
brane\cite{low-AD-problem}. 
One can solve this problem by introducing non-trivial defect
configuration structures\cite{low-AD-solution_matsuda}.
Moreover, there are more arguments
about the mechanism of baryogenesis with low-scale inflationary
scale\cite{low_baryo}.
Although there is no ultimate solution to this problem,
it is not incorrect to expect that baryogenesis could be successful even
if the inflationary scale is as low as the TeV scale.

\section{Conclusions and Discussions}
\hspace*{\parindent}
A new mechanism for generating the curvature perturbation at
 the end of inflation is discussed in this paper.
The dominant contribution to the primordial curvature perturbation
may be generated by this new mechanism, which converts isocurvature
perturbation of a light field into curvature perturbation during
the period of instant preheating.
The light field is ``not'' the inflaton field.
Our new mechanism is important in inflationary models where the kinetic
energy is significant at the time of reheating\footnote{A conserved
angular momentum may be significant in some other cases\cite{Lefteris,
Grandi}.}. 
What we have presented in this paper is a new mechanism for
 generating the curvature perturbation at the end of inflation, which
 works even if (1)the kinetic energy is significant, or 
  (2)the inflation potential is not a hybrid-type.
Thus, our present model complements the original
scenarios\cite{delta-N-Lyth,  matsuda_elliptic,
matsuda_ellip2, SSB-curvaton}  to give the 
paradigm of generating the curvature perturbation at the end of
inflation. 

\section{Acknowledgment}
We wish to thank K.Shima for encouragement, and our colleagues at
Tokyo University for their kind hospitality.

\appendix
\section{Instant preheating in two-field models}
The key equation in this paper is eq.(\ref{n_chi}), which describes the
number density of particles produced by instant preheating. 
However, the original result obtained in ref.\cite{inst_preheating} was
\begin{equation}
n_\chi = \frac{(g|\dot{\phi_1}(t_*)|)^{3/2}}{8\pi^3}
\exp\left[-\frac{\pi m_\chi^2}{g |\dot{\phi_1}(t_*)|}
\right],
\end{equation}
which applies only to a single-field model.
Moreover, in a two-field model there must be some restriction on the
assumed value of $\phi_2$ and its potential, which was not explicit in
the above arguments.
Therefore, in this appendix we will present in more detail the
derivation of eq.(\ref{n_chi}) and also some restriction on $\phi_2$ and
its potential, which applies to a two-field model. 
In some cases one may find sensible amount of 
non-gaussianity\cite{paper-non-gaussianity}. 
A standard adiabatic curvature perturbation generated by the fluctuation
in the inflaton field $\phi_1$ is also important if it is generated during
inflation.
As we will discuss later in this appendix, this condition puts some
restriction on the inflaton potential. 

Let us first consider what happens if both $\phi_1$ and $\phi_2$ are
dynamical at the same time. 
Here we will assume that $\phi_1$ is heavier than $\phi_2$, and the
$\phi_1$-oscillation starts when $H\sim m_1$.
If the condition $\dot{\phi_1}> \pi g|\phi_2|^2$ is satisfied at the
bottom of the $\phi_1$-potential, 
the instant preheating terminates the 
$\phi_1$-oscillation before $\phi_2$ starts to oscillate.
In this case one may assume that the value of $\phi_2$ is a constant
during the instant preheating.
We have considered this situation in this paper.
On the other hand, if the above condition is not satisfied for some time
during the $\phi_1$-oscillation, the
$\phi_1$-oscillation continues until it loses its energy.
The Hubble constant decreases with time as the amplitude
of the $\phi_1$-oscillation decreases, and finally $\phi_2$ starts to
oscillate when $H\sim m_2$.
Here we have assumed that the decay constant $\Gamma_{\phi_1}$ is much
smaller than $m_2$, otherwise the instant preheating cannot play
significant role in reheating.
In this peculiar case, both $\dot{\phi_1}$ and $\dot{\phi_2}$ are 
equally important.
The most obvious example in this direction is discussed by Kolb
et. al.\cite{SSB-curvaton} who considered a model with a weakly broken
$U(1)$ symmetry and 
presented a mechanism to generate curvature perturbations after 
the end of the slow-roll phase of inflation.\footnote{If the $U(1)$
symmetry is strongly broken, the trajectory becomes a 
Lissajous curve. 
The instant preheating occurs when 
the trajectory satisfies the condition
$|\dot{\phi}(t_*)| <g|\phi(t_*)|^2$ at a minimum distance 
$\phi(t_*)$, where $\phi$ is
defined as $\phi^2 \equiv \phi_1^2 + \phi_2^2$.
Thus, in the case where the $U(1)$ symmetry is strongly broken, and also
if the motion of 
$\phi_2$ is not negligible, the calculation of the
curvature perturbation requires numerical method.
This case may be interesting, but it is beyond the scope of this paper.}
Let us make a short review of this scenario to complement our
discussions in this paper.
Here we assume that the potential takes the simple form
\begin{equation}
\label{ssb-pot}
V(\phi_1,\phi_2)=\frac{m^2}{2}\left[\phi_1^2 + 
\frac{\phi_2^2}{(1+x)}\right],
\end{equation}
where $x$ is a small constant that measures the symmetry breaking.
Given the potential (\ref{ssb-pot}), the solution for the background
dynamics can be computed neglecting the expansion of the Universe.
We can neglect the expansion of the Universe since after a first
transient the trajectory in field space will be mostly along the radial
direction, and also the instant preheating occurs immediately.
Introducing a complex field $\phi \equiv \phi_1 + i \phi_2
=|\phi|e^{i\theta}$, the background field dynamics is then given by
\begin{eqnarray}
\phi_1(t) &=&|\phi_0| \cos \theta_0 \cos mt \nonumber\\
\phi_2(t)&=&|\phi_0| \sin \theta_0 \cos \left(\frac{mt}{1+x}\right),
\end{eqnarray}
where the index $0$ is for the initial conditions.
Linearizing the trajectory\cite{SSB-curvaton}, one can obtain a minimum
distance $|\phi(t_*)|$,
\begin{equation}
|\phi(t_*)| = \frac{|\phi_0|\pi x}{2\sqrt{2}}|\sin 2\theta_0|.
\end{equation}
After a simple calculation\cite{SSB-curvaton}, the velocity of the field
is obtained as 
\begin{equation}
\dot{\phi}(t_*)\simeq m |\phi_0|\left(1-x\sin^2 \theta_0\right)^{1/2}.
\end{equation}
The preheat particles $\chi$ generated will be characterized by an
effective mass $m_\chi= g |\phi_*|$.
Hence, the comoving number density of $\chi$ particles produced by the
instant preheating is given by 
\begin{equation}
\label{SSB-chi}
n_\chi = \frac{(g|\dot{\phi}(t_*)|)^{3/2}}{8\pi^3}
\exp\left[-\frac{\pi g |\phi(t_*)|^2}{|\dot{\phi}(t_*)|}
\right].
\end{equation}
If $\phi_2$ is so light that it cannot move before the instant
preheating, the trajectory of the oscillation becomes linear and 
one obtains eq.(\ref{n_chi}) instead of eq.(\ref{SSB-chi}).

Although in our mechanism slow-roll inflation is not required for
generating curvature perturbations, it is important to consider a
constraint that the curvature perturbation generated at the end of
inflation should dominate over the standard inflaton perturbation at
horizon exit. 
Fast-roll inflation is an obvious example, where the inflaton field has
a mass of $O(H)$.
Since the inflaton is not light in fast-roll inflation, one may simply 
neglect the standard inflaton perturbation.
On the other hand, the standard inflaton perturbation may be
important if the large number of e-foldings is generated due
to a ``slow-roll'' inflaton.
The constraint that the curvature perturbations generated by an instant
preheating, which are given by eq.(\ref{zeta-1}), eq.(\ref{zeta-2}) or
eq.(\ref{zeta-PR-DBI}), should dominate over the standard inflaton
perturbation $\zeta_{st}$ is written as $\zeta/\zeta_{st} \gg 1$.
This constraint depends crucially on the scenario of inflation, and
requires further study.

In this paper we have assumed that the impact parameter $\phi_2$ is much
lager than its fluctuation $\delta \phi_2$ that exits horizon during
inflation. 
This condition is very crucial to obtain the required
perturbation spectrum.
Once this condition is violated, higher order terms that have been
neglected in the above calculation may become significant
giving rise to a large non-gaussianity\cite{paper-non-gaussianity}.
Let us see more details.
The expression for $\delta n_\chi / n_\chi$, which was given by
eq.(\ref{1storder}), has the additional terms proportional to $(\delta
\phi_2)^2$, which are given by
\begin{equation}
\label{2ndorder}
\frac{\delta n_\chi}{n_\chi} \simeq
-\frac{2\pi g |\phi_2(t_*)|^2}{|\dot{\phi}_1(t_*)|} 
\frac{|\delta \phi_2(t_*)|}{|\phi_2(t_*)|}
-\frac{1}{2}\left(\frac{2\pi g|\phi_2(t_*)|^2}{|\dot{\phi}_1(t_*)|}
-\frac{4\pi^2 g^2 |\phi_2(t_*)|^4}{|\dot{\phi}_1(t_*)|^2}
\right)\frac{|\delta \phi_2(t_*)|^2}{|\phi_2(t_*)|^2}.
\end{equation}
The non-gaussianity parameter, if observable, is
\begin{equation}
-\frac{3}{5}f_{NL} = 
\frac{|\dot{\phi}_1(t_*)|}{4\pi g \alpha|\phi_2(t_*)|^2} 
-\frac{1}{2\alpha}.
\end{equation}
Considering a condition 
$m_1^2 |\phi_1(t_i)|^2 \simeq |\dot{\phi}_1(t_*)|^2$,
non-gaussianity becomes obviously large when
 $\phi_2(t_*)<m_1 <\delta \phi_2(t_*)$.


\begin{thebibliography}{1}
\bibitem{Books-EU}
See the following books and the references therein.
E. W. Kolb and M. S. Turner,
{\it  The Early Universe,  Addison-Wesley, US, (1990)};
A. R. Liddle and D. H. Lyth, 
{\it Cosmological inflation and large-scale structure,
Cambridge, UK: Univ. Pr. (2000)}.
\bibitem{Lyth_constraint}
D. H. Lyth,
{\it Can the curvaton paradigm accommodate a low inflation scale?,
	Phys.Lett.B579(2004)239} [hep-th/0308110].
\bibitem{MSSM-infla}
R. Allahverdi, K. Enqvist, J. Garcia-Bellido and A. Mazumdar,
{Gauge invariant MSSM inflaton} [hep-ph/0605035];
R. Allahverdi, A. Kusenko and A. Mazumdar,
{\it A-term inflation and the smallness of neutrino
	masses}[hep-ph/0608138].
\bibitem{curvaton_1}
S. Mollerach, 
{\it Isocurvature baryon perturbations and inflation,
	Phys.Rev.D42(1990)313};
A. D. Linde, V. Mukhanov,
{\it Nongaussian Isocurvature Perturbations from Inflation,
	Phys.Rev.D56(1997)535} [astro-ph/9610219];
K. Enqvist, M. S. Sloth,
{\it Adiabatic CMB perturbations in pre-big bang string cosmology,
	Nucl.Phys.B626(2002)395} [hep-ph/0109214].
\bibitem{curvaton_2}
T. Moroi, T. Takahashi,
{\it Effects of Cosmological Moduli Fields on Cosmic Microwave
	Background, Phys.Lett.B522(2001)215} [hep-ph/0110096]
D. H. Lyth, D. Wands,
{\it Generating the curvature perturbation without an inflaton,
	Phys.Lett.B524(2002)5} [hep-ph/0110002].
\bibitem{curvaton_liberate}
K. Dimopoulos, D. H. Lyth, 
{\it Models of inflation liberated by the curvaton hypothesis,
	Phys.Rev.D69(2004)123509,2004} [hep-ph/0209180];
M. Giovannini,
{\it Tracking curvaton(s)?, Phys.Rev.D67(2003)123512} [hep-ph/0310024].
\bibitem{low_inflation}
P. Kanti and K. A. Olive, 
{\it Assisted chaotic inflation in higher dimensional theories,
Phys. Lett. B464(1999)192} [hep-ph/9906331];
N. Arkani-Hamed, S. Dimopoulos, N. Kaloper, and J. March-Russell,
{\it Rapid asymmetric inflation and early cosmology in theories with
submillimeter dimensions, Nucl.Phys.B567(2000)189} [hep-ph/9903224];
R. N. Mohapatra, A. Perez-Lorenzana, and C. A. de S. Pires,
{\it Inflation in models with large extra dimensions driven by a bulk
scalar field, Phys.Rev.D62(2000)105030} [hep-ph/0003089];
A. Mazumdar, S. Panda and A. Perez-Lorenzana,
{\it Assisted inflation via tachyon condensation,
	Nucl.Phys.B614(2001)101} [hep-ph/0107058];
A. Mazumdar,
{\it Extra dimensions and inflation, Phys.Lett.B469(1999)55}
	[hep-ph/9902381];
A. M. Green and A. Mazumdar, 
{\it Dynamics of a large extra dimension inspired hybrid inflation model,
Phys.Rev.D65(2002)105022} [hep-ph/0201209];
T. Matsuda,
{\it Nontachyonic brane inflation, Phys.Rev.D67(2003)083519}
[hep-ph/0302035];
T. Matsuda,
{\it F term, D term and hybrid brane inflation, JCAP 0311(2003)003}
[hep-ph/0302078];
T. Matsuda,
{\it Successful D term inflation with moduli, Phys.Lett.B423(1998)35}
[hep-ph/9705448].
\bibitem{matsuda_defectinfla}
T. Matsuda,
{\it Topological hybrid inflation in brane world, JCAP 0306(2003)007}
[hep-ph/0302204];
T. Matsuda,
{\it Q ball inflation, Phys.Rev.D68(2003)127302} [hep-ph/0309339];
T. Matsuda,
{\it Brane Q Ball, branonium and brane Q ball inflation} [hep-ph/0402223].
\bibitem{matsuda_curvaton}
T. Matsuda,
{\it Curvaton paradigm can accommodate multiple low inflation scales,
Class.Quant.Grav.21(2004)L} [hep-ph/0312058].
\bibitem{curvaton_added}
 K. Dimopoulos and G. Lazarides,
{\it Modular inflation and the orthogonal axion as curvaton,
  Phys.Rev.D73(2006)023525} [arXiv:hep-ph/0511310];
  K. Dimopoulos and G. Lazarides,
{\it Modular inflation and the orthogonal axion as curvaton,
 Phys.Rev.D73(2006)023525}[arXiv:hep-ph/0511310];
  K. Dimopoulos,
{\it Inflation at the TeV scale with a PNGB curvaton,
	Phys.Lett.B634(2006)331} [arXiv:hep-th/0511268];
  K.~Dimopoulos,
{\it Inflation at the TeV scale with a PNGB curvaton,
  Phys.Lett.B634(2006)331}[arXiv:hep-th/0511268].
\bibitem{delta-N-Lyth}
D. H. Lyth,
{\it Generating the curvature perturbation at the end of inflation,
JCAP 0511:006,2005} [astro-ph/0510443].
\bibitem{alternate}
F. Bernardeau, L. Kofman and J.P. Uzan, 
{\it Modulated fluctuations from hybrid inflation,
	Phys.Rev.D70(2004)083004} [astro-ph/0403315];
K. Enqvist, A. Mazumdar and M. Postma,
{\it Challenges in generating density perturbations from a 
fluctuating inflaton coupling, Phys.Rev.D67(2003)121303}
	[astro-ph/0304187];
A. Mazumdar and M. Postma,
{\it Evolution of primordial perturbations and a fluctuating decay 
rate, Phys.Lett.B573(2003), [Erratum-ibid.\ B585(2004)295]}
[arXiv:astro-ph/0306509].
\bibitem{matsuda_elliptic}
T. Matsuda,
{\it Elliptic Inflation: Generating the curvature perturbation without
	slow-roll, JCAP 0609:003,2006} [hep-ph/0606137].
\bibitem{matsuda_ellip2}
T. Matsuda,
{\it Brane inflation without slow-roll} [astro-ph/0610402]. 
\bibitem{Riotto-Lyth}
D. H. Lyth and A. Riotto,
{\it Generating the Curvature Perturbation at the End of Inflation in
	String Theory} [astro-ph/0607326].
\bibitem{KS-throat}
R. Klebanov and Matthew J. Strassler,
{\it Supergravity and a confining gauge theory: Duality cascades and chi
	SB resolution of naked singularities, JHEP 0008:052,2000}
[hep-th/0007191].
\bibitem{KK-string_Kofman}
L. Kofman and  P. Yi, 
{\it Reheating the universe after string theory inflation,
Phys.Rev.D72:106001,2005} [hep-th/0507257].
\bibitem{inst_preheating}
G. N. Felder, L. Kofman and Andrei D. Linde,
{\it Instant preheating, Phys.Rev.D59:123523,1999}[hep-ph/9812289].
\bibitem{Tong-DBI}
M. Alishahiha, E. Silverstein and D. Tong,
{\it DBI in the sky, Phys.Rev.D70:123505,2004}[hep-th/0404084].
\bibitem{beauty_is}
L. Kofman, A. Linde, X. Liu, A. Maloney, L. McAllister and
 Eva Silverstein,
{\it Beauty is attractive: Moduli trapping at enhanced symmetry points,
JHEP 0405:030,2004}[hep-th/0403001].
\bibitem{ultra-motion}
L. McAllister and I. Mitra,
{\it Relativistic D-brane scattering is extremely inelastic,
JHEP 0502:019,2005} [hep-th/0408085].
\bibitem{SSB-curvaton}
W. Kolb, A. Riotto and A. Vallinotto,
{\it Curvature perturbations from broken symmetries,
Phys.Rev.D71:043513,2005} [astro-ph/0410546];
W. Kolb, A. Riotto and A. Vallinotto,
{\it Non-gaussianity from broken symmetries, Phys.Rev.D73:023522,2006}
	[astro-ph/0511198].
\bibitem{TeV-hybrid}
D. H. Lyth,
{\it Constraints on TeV scale hybrid inflation and comments on nonhybrid
	alternatives, Phys.Lett.B466:85-94,1999} [hep-ph/9908219].
\bibitem{matsuda_KK}
T. Matsuda,
{\it Kaluza-Klein modes in hybrid inflation, Phys.Rev.D66(2002)107301}
[hep-ph/0209214].
\bibitem{warped-reaheating}
D. Chialva, G. Shiu and B. Underwood, 
{\it Warped reheating in multi-throat brane inflation,
JHEP 0601:014,2006} [hep-th/0508229];
X. Chen and  S.-H.Henry Tye,
{\it Heating in brane inflation and hidden dark matter,
JCAP 0606:011,2006} [hep-th/0602136].
\bibitem{string_relic}
T. Damour and A. Vilenkin, 
{\it Gravitational radiation from cosmic (super)strings: Bursts,
	stochastic background, and observational windows,
Phys.Rev.D71:063510,2005} [hep-th/0410222];
T. Matsuda,
{\it String production after angled brane inflation,
	Phys.Rev.D70(2004)023502} [hep-ph/0403092].
\bibitem{matsuda_brane_monopolesandwalls}
T. Matsuda, 
{\it Formation of monopoles and domain walls after brane
	inflation, JHEP 0410(2004)042} [hep-ph/0406064].
T.Matsuda,
{\it Formation of cosmological brane defects}
[hep-ph/0402232];
T. Matsuda,
{\it Incidental Brane Defects, JHEP 0309(2003)064} [hep-th/0309266];
\bibitem{matsuda_brane_relics}
T. Matsuda,
{\it Brane necklaces and brane coils,
JHEP 0505(2005)015} [hep-ph/0412290];
T.Matsuda,
{\it Primordial black holes from cosmic necklaces, JHEP 0604:017,2006}
 [hep-ph/0509062]
T. Matsuda,
{\it Dark matter production from cosmic necklaces, JCAP 0604:005,2006}
 [hep-ph/0509064];
A. Avgoustidis and E.P.S. Shellard,
{\it Cycloops: Dark matter or a monopole problem for brane inflation?, 
JHEP 0508:092,2005} [hep-ph/0504049];
T. Matsuda,
{\it Primordial black holes from monopoles connected by strings}
	[hep-ph/0509061]. 
\bibitem{matsuda_wall}
T. Matsuda,
{\it Weak scale inflation and unstable domain walls,
Phys.Lett.B486(2000)300} [hep-ph/0002194];
T. Matsuda,
{\it On cosmological domain wall problem in supersymmetry models,
Phys.Lett.B436(1998)264} [hep-ph/9804409].

\bibitem{low-x-decay-baryo}
A. Masiero, M. Peloso, L. Sorbo, and R. Tabbash, 
{\it Baryogenesis versus proton stability in theories with extra
	dimensions, 
Phys.Rev.D62(2000)063515} [hep-ph/0003312].
\bibitem{low-x-decay-baryo_matsuda}
T. Matsuda,
{\it Baryon number violation, baryogenesis and defects with extra
dimensions, Phys.Rev.D66(2002)023508} [hep-ph/0204307];
T. Matsuda,
{\it Enhanced baryon number violation due to cosmological defects with
localized fermions along extra dimensions, Phys.Rev.D65(2002)107302}
[hep-ph/0202258].
\bibitem{AD}
I. Affleck and M. Dine,
{A new mechanism for baryogenesis, Nucl.Phys.B249(1985)361}.
\bibitem{low-AD-problem}
R. Allahverdi, K. Enqvist, A. Mazumdar and A. Perez-Lorenzana,
{\it Baryogenesis in theories with large extra spatial dimensions,
Nucl.Phys. B618(2001)277} [hep-ph/0108225].
\bibitem{low-AD-solution_matsuda}
T. Matsuda,
{\it Affleck-Dine baryogenesis in the local domain,
Phys.Rev.D65(2002)103502} [hep-ph/0202211];
T. Matsuda,
{\it Hybridized Affleck-Dine baryogenesis, Phys.Rev.D67(2003)127302}
[hep-ph/0303132];
T. Matsuda,
{\it Affleck-Dine baryogenesis after thermal brane inflation,
Phys.Rev.D65(2002)103501} [hep-ph/0202209].
\bibitem{low_baryo}
G. R. Dvali, G. Gabadadze, 
{\it Nonconservation of global charges in the brane universe and
	baryogenesis,e 
Phys.Lett.B460(1999)47} [hep-ph/9904221];
A. Pilaftsis, {\it Leptogenesis in theories with large extra dimensions,
Phys.Rev.D60(1999)105023} [hep-ph/9906265];
S. Davidson, M. Losada, and A. Riotto,
{\it A new perspective on baryogenesis, Phys.Rev.Lett.84(2000)4284}
[hep-ph/0001301];
T. Matsuda,
{\it Activated sphalerons and large extra dimensions,
Phys.Rev.D66(2002)047301} [hep-ph/0205331];
T. Matsuda,
{\it Electroweak baryogenesis mediated by locally supersymmetry breaking
defects, Phys.Rev.D64(2001)083512} [hep-ph/0107314].
T. Matsuda,
{\it Defect mediated electroweak baryogenesis and hierarchy,
J.Phys.G27(2001)L103} [hep-ph/0102040].
\bibitem{paper-non-gaussianity}
N. Bartolo, S. Matarrese and A. Riotto,
{\it Non-Gaussianity from inflation, Phys.Rev.D65(2002)103505}
[hep-ph/0112261];
N. Barnaby and J. M. Cline,
{\it Nongaussianity from Tachyonic Preheating in Hybrid Inflation} 
[astro-ph/0611750];
N. Barnaby and J. M. Cline,
{\it Nongaussian and nonscale-invariant perturbations from tachyonic
preheating in hybrid inflation, Phys.Rev.D73(2006)106012} 
[astro-ph/0601481].
\bibitem{Lefteris}
E. Papantonopoulos and V. Zamarias,
{\it AdS/CFT Correspondence and the Reheating of the Brane-Universe, 
JHEP 0410, 051 (2004)} [hep-th/0408227]
\bibitem{Grandi}
C. Germani, N.E. Grandi and A. Kehagias,
{\it A Stringy Alternative to Inflation: The Cosmological Slingshot
Scenario} [hep-th/0611246]
\end{thebibliography}
\end{document}